\documentclass[twoside,twocolumn,9pt]{article}

\usepackage[utf8]{inputenc}
\usepackage{subfig}
\usepackage{t1enc,latexsym,amsfonts,amssymb,amsmath}
\usepackage{bm}
\usepackage{graphics}
\usepackage{wasysym}
\usepackage{setspace}
\usepackage{url}
\usepackage{nicefrac}

\usepackage{extsizes}
\usepackage[super,sort&compress,comma]{natbib}
\usepackage[version=3]{mhchem}
\usepackage[left=1.5cm, right=1.5cm, top=1.785cm, bottom=2.0cm]{geometry}
\usepackage{balance}
\usepackage{times,mathptmx}
\usepackage{sectsty}
\usepackage{graphicx}

\usepackage{lastpage}
\usepackage[format=plain,justification=justified,singlelinecheck=false,font={stretch=1.125,small,sf},labelfont=bf,labelsep=space]{caption}
\usepackage{float}
\usepackage{fancyhdr}
\usepackage{fnpos}
\usepackage[english]{babel}
\usepackage{array}
\usepackage{droidsans}
\usepackage{charter}
\usepackage[T1]{fontenc}
\usepackage[usenames,dvipsnames]{xcolor}
\usepackage{setspace}
\usepackage[compact]{titlesec}
\usepackage{color}

\usepackage{blindtext}

\usepackage{epstopdf}

\usepackage{dcolumn}   

\definecolor{cream}{RGB}{222,217,201}

\begin{document}

\pagestyle{fancy}
\thispagestyle{plain}
\fancypagestyle{plain}{

\renewcommand{\headrulewidth}{0pt}
}

\makeFNbottom
\makeatletter
\renewcommand\LARGE{\@setfontsize\LARGE{15pt}{17}}
\renewcommand\Large{\@setfontsize\Large{12pt}{14}}
\renewcommand\large{\@setfontsize\large{10pt}{12}}
\renewcommand\footnotesize{\@setfontsize\footnotesize{7pt}{10}}
\makeatother

\renewcommand{\thefootnote}{\fnsymbol{footnote}}
\renewcommand\footnoterule{\vspace*{1pt}%
\color{cream}\hrule width 3.5in height 0.4pt \color{black}\vspace*{5pt}}
\setcounter{secnumdepth}{5}

\makeatletter
\renewcommand\@biblabel[1]{#1}
\renewcommand\@makefntext[1]%
{\noindent\makebox[0pt][r]{\@thefnmark\,}#1}
\makeatother
\renewcommand{\figurename}{\small{Fig.}~}
\sectionfont{\sffamily\Large}
\subsectionfont{\normalsize}
\subsubsectionfont{\bf}
\setstretch{1.125} 
\setlength{\skip\footins}{0.8cm}
\setlength{\footnotesep}{0.25cm}
\setlength{\jot}{10pt}
\titlespacing*{\section}{0pt}{4pt}{4pt}
\titlespacing*{\subsection}{0pt}{15pt}{1pt}

\fancyfoot{}
\fancyfoot[RO]{\footnotesize{\sffamily{1--\pageref{LastPage} ~\textbar  \hspace{2pt}\thepage}}}
\fancyfoot[LE]{\footnotesize{\sffamily{\thepage~\textbar\hspace{3.45cm} 1--\pageref{LastPage}}}}
\fancyhead{}
\renewcommand{\headrulewidth}{0pt}
\renewcommand{\footrulewidth}{0pt}
\setlength{\arrayrulewidth}{1pt}
\setlength{\columnsep}{6.5mm}
\setlength\bibsep{1pt}

\makeatletter
\newlength{\figrulesep}
\setlength{\figrulesep}{0.5\textfloatsep}

\newcommand{\topfigrule}{\vspace*{-1pt}%
\noindent{\color{cream}\rule[-\figrulesep]{\columnwidth}{1.5pt}} }

\newcommand{\botfigrule}{\vspace*{-2pt}%
\noindent{\color{cream}\rule[\figrulesep]{\columnwidth}{1.5pt}} }

\newcommand{\dblfigrule}{\vspace*{-1pt}%
\noindent{\color{cream}\rule[-\figrulesep]{\textwidth}{1.5pt}} }

\makeatother


\newcommand{\argc}[1]{\left[#1\right]}
\newcommand{\arga}[1]{\left\lbrace #1\right\rbrace }
\newcommand{\argp}[1]{\left(#1\right)}
\newcommand{\valabs}[1]{\vert #1\vert}
\newcommand{\moy}[1]{\left\langle  #1 \right\rangle }
\newcommand{\moydes}[1]{\overline{#1}}
\renewcommand{\vec}[1]{{\bf #1}}
\newcommand{\meevid}[1]{\textcolor{VioletRed}{#1}}
\newcommand{\KM}[1]{\textcolor{blue}{Comment KM: #1}}
\newcommand{\comment}[1]{\textcolor{OliveGreen}{#1}}



\twocolumn[
  \begin{@twocolumnfalse}
\vspace{3cm}
\sffamily
\begin{tabular}{m{4.5cm} p{13.5cm} }

& \noindent\LARGE{\textbf{Precursors of fluidisation in the creep response of a soft glass}} \\

 & \noindent\large{Raffaela Cabriolu$^{\ast}$\textit{$^{a}$}, J\"urgen Horbach\textit{$^{b}$}, Pinaki Chaudhuri\textit{$^{c}$} and Kirsten Martens\textit{$^{d}$}} \\
& \noindent\normalsize{
Using extensive numerical simulations, we study the the fluidisation process of dense amorphous materials subjected to an external shear stress, using a three-dimensional colloidal glass model.
%
%
In order to disentangle possible boundary effects from finite size effects in the process of fluidisation, we implement a novel geometry-constrained protocol with periodic boundary conditions.
We show that this protocol is well controlled and that the long time fluidisation dynamics is, to a great extent, independent of the details of the protocol parameters. 
Our protocol therefore provides an ideal tool to investigate the bulk dynamics prior to yielding and to study finite size effects regarding the fluidisation process. 
Our study reveals the existence of precursors to fluidisation observed as a peak in the strain-rate fluctuations, that allows for a robust definition of a fluidisation time.
Although the exponents in the power-law creep dynamics seem not to depend significantly on the system size, we reveal strong finite size effects for the onset of fluidisation.
} \\

\end{tabular}

 \end{@twocolumnfalse} \vspace{0.6cm}

  ]
  


\renewcommand*\rmdefault{bch}\normalfont\upshape
\rmfamily
\section*{}
\vspace{-1cm}


\footnotetext{\textit{$^{*}$~E-mail:} raffaela.cabriolu@ntnu.no}
\footnotetext{\textit{$^{a}$~Department of Chemistry, Norwegian University of Science and Technology (NTNU), H\o gskoleringen 5, 7491 Trondheim, Norway.}}
\footnotetext{\textit{$^{b}$~The Institute of Mathematical Sciences, Taramani, Chennai 600113, India.}}
\footnotetext{\textit{$^{c}$~Institut f\"ur Theoretische Physik II, Heinrich-Heine-Universit\"at D\"usseldorf, 40225 D\"usseldorf, Germany.}}
\footnotetext{\textit{$^{d}$~Univ.~Grenoble Alpes, CNRS, LIPhy, 38000 Grenoble, France.}}




\section{Introduction}
\label{section-intro}

Yield stress materials \cite{bonn2009yield, zapperi2012current},
although ubiquitous in everyday life, remain a challenging topic
both for a fundamental physical understanding as well as for applied
engineering problems. These materials are characterised by the
feature that they behave solid-like for a small externally applied
force, but can yield above a given yield stress either towards
failure or towards a steady flow, depending on the brittleness of
the material \cite{ozawa2018random, popovic2018elasto}. One way to
probe the complex response of these materials is to apply stresses,
close to this yielding threshold, and study the transient dynamics
preceding the yielding. Although macroscopic features of the yielding
have been studied extensively for a long time, it is only recently
that molecular dynamics simulations and scattering experiments have
opened the way for a more microscopic understanding of the underlying
processes, that lead to failure and flow, in such circumstances.
This motivates our investigation of the creep dynamics of disordered
materials using a microscopic approach, in order to reveal the
physics involved and to search for possible signs of precursors.
In this context, predicting the moment at which a material is going
to yield is of course a very important goal, especially with respect
to technical applications.

Many of these materials respond to an externally applied external
load in the form of slow creep dynamics that can be succeeded either
by complete arrest, catastrophic failure or rapid fluidisation.
This observation is not only true for crystalline and amorphous
hard solids, e.g. metallic systems, but has been reported as well
in soft materials, such as dense emulsions and gels. In this context
the measured macroscopic quantities of interest are usually the
time-dependent deformation rate $\dot{\gamma}(t)$ and the fluidisation
or failure time $\tau_{f}$ \cite{skrzeszewska2010fracture,Divoux2011}.
It is only relatively recent that the particular processes inducing
the macroscopic creep dynamics have been also addressed on a more
microscopic level. Experimentally this is possible for example by
revealing the time evolution of spatially resolved plastic activity
using dynamic light scattering techniques \cite{pons2016spatial}
or confocal microscopy \cite{sentjabrskaja2015creep}. Another
extremely insightful microscopic approach is of course the investigation
of the creep and fluidisation using particle based simulations,
which has been used both for dense particle systems \cite{Chaudhuri2013}
and for network-forming gels \cite{landrum2016delayed}.

The occurrence of a yield stress implies the existence of a rigid
solid. This rigid phase, however, does not correspond to an equilibrium
state in the thermodynamic limit where the free energy of any phase
of a system cannot depend on the shape of its boundary
\cite{lebowitz1968statistical, ruelle1999statistical}. Thus, solids
that sustain a finite stress are in a non-equilibrium state and,
provided that one waits long enough, they would eventually evolve
into a stress-free state under the application of even an infinitesimal
stress \cite{penrose2002statistical, sausset2010solids, saw2016rigidity,
saw2016nonaffine, nath2018existence}.  The time associated with the
approach of the latter equilibrium state may in general exceed any
observable scales and, in this sense, rigid or yield-stress materials
are very long-lived metastable systems. The transformation from a
rigid solid to a stress-free solid in the limit of zero strain rate,
$\dot{\gamma} \to 0$, is not well understood and in most cases this
limit has to be inferred from an extrapolation from finite strain
rates.  Only in the case of a two-dimensional crystal, a recent
computer simulation study \cite{nath2018existence} demonstrated
that the transformation from a metastable rigid crystal into a
stress-free solid at finite strain rate are kinetic processes,
associated with an underlying first-order phase transition at
$\dot{\gamma}=0$ and zero deformation. An example for such a kinetic
process in crystalline materials is known as Andrade creep
\cite{miguel2002dislocation}, reflected by a power law of the strain
rate as a function of the applied stress with an exponent close to
$2/3$. Of course, it remains to be shown how Andrade creep is exactly
linked to the phase-ordering kinetics, thus being an analogue to
spinodal decomposition in usual first-order transformations, such
as liquid-gas or liquid-solid transitions.

In amorphous solids, the appearance of rigidity is due to a probable
shift of the Newtonian regime to inaccessible long time scales,
and there is not a similar underlying
first-order phase transition in the limit
$\dot{\gamma} \to 0$ as for the crystalline case  \cite{nath2018existence}.  Meanwhile, experiments
and simulations studying creep in amorphous systems have found a
variety of power-law exponents for the decay of the deformation
rate, ranging between $\nicefrac{1}{3}$ and $1.0$ (the latter value
corresponding to logarithmic creep), with a multitude of values
in-between \cite{bauer2006collective,
Divoux2011a,Leocmach2014,ballesta2016creep,Chaudhuri2013,sentjabrskaja2015creep,landrum2016delayed,
liu2018mean} (see also the extensive reviews of Bonn et
al.~\cite{bonn2017yield} and Nicolas et al.~\cite{nicolas2017deformation}).
Thus, in the context of amorphous systems, the creep response shows
a multitude of non-universal dependencies, notably on the preparation
protocol prior to the application of the step stress (quench or
pre-shear), on temperature, age and also on the dominant microscopic
processes at play during the creep regime.  In some network forming
gel systems, the initial creep regime has been shown to be completely
reversible and one expects the power-law creep to be a result of
visco-elastic effects in a fractal gel network \cite{jaishankar2013power,
aime2018power}.  On the other hand, there have been studies on the
basis of molecular dynamics simulations explaining that the power-law
creep in a variety of glassy systems can be related to a percolation
of mobile regions and thus plasticity \cite{shrivastav2016yielding}.

In a broader context, understanding and characterizing the physical
mechanisms leading to observed rheological response of the amorphous
materials in the vicinity of the yield stress threshold is of
fundamental interest, with predictions of a dynamical transition
at the threshold value.  Therefore, the investigation of possible
critical behaviour near the yield stress along with associated
possible finite size effects, has been subject of many studies
\cite{salerno2012avalanches,talamali2011avalanches,lin2014scaling,
liu2016driving} but remains still a debated topic.  One should
note that some experiments have reported size effects in the measured
value of threshold.  In this context, a stress-controlled protocol
at finite temperature allows for exploration of the creep response
both above and below the threshold, as well as an investigation of
possible precursors before complete fluidisation. Further, the idea
of the underlying critical transition also motivates scrutiny of
how system size influences the observed behaviour.  For the case
of applied shear rate, there have been diverse studies, looking at
finite size effects, both in transient response and steady-state
behaviour, but not much has been checked for applied stress situations.

One of the quantities of interest in the case of transient response,
is the time-scale for fluidisation (or failure), $\tau_{f}$, starting
from an amorphous material at rest, and the results also vary and
basically fall in two classes. Works on athermal creep in dense
amorphous materials usually find power law scalings of the fluidisation
time \cite{Divoux2011a, liu2018mean} whereas thermally activated
processes lead rather to an inverse exponential dependence
\cite{gopalakrishnan2007delayed,gibaud2010heterogeneous,Lindstrom2012,merabia2016thermally}.

In this work we concentrate on thermally induced non-reversible
creep in dense disordered materials.  Using an unconventional
protocol combining a geometry imposed shear stress with periodic
boundary conditions, we are able to disentangle finite size effects
from boundary effects in the creep and fluidisation dynamics.

The article is structured as follows: In a first section we describe
in detail the particle model, its rheological features and the novel
protocol. We check in the second section for the robustness of this
protocol with respect to the involved control parameters. Using
extensive simulations averaging over a large number of realisations
for different system sizes we then study the compliance curves as
a function of the imposed shear stress which reveals the onset of
plasticity in the transient response. The main result of our work
concerns the study of the bulk dynamics prior to fluidisation. We
show that the onset of fluidisation is accompanied by a maximum in
the strain rate fluctuations, which one can interpret as a precursor
to yielding. This feature allows to define in an unambiguous manner
a fluidisation criterion and we find that the corresponding
fluidisation time occurs at a given yield strain, which shows strong
finite size effects. The last section is reserved for a detailed
discussion of our results and the presentation of some ideas for
potentially interesting future works.

\section{Model and Scheme}

For our study, we consider a three dimensional glass-forming 50:50 colloidal binary mixture, 
introduced in an earlier work by Zausch et al.~\cite{Zausch2009}. In this model, the colloidal particles,
of species a and b, interact via the pair-wise Yukawa potential: 
\begin{equation}
U_{1,2}(r) = \epsilon_{1,2} d_{1,2} \frac{ exp(-k_{1,2}(r-d_{1,2}))} {r} \quad 1,2 \in \{a,b\}
\end{equation}
where, $\epsilon$ is the energy unit, $d$ the diameter of the
colloidal particle, $k$ the screening length and $r$ the distance
between the two interacting particles.  As in Ref. ~\cite{Zausch2009},
the parameters are chosen as: $\epsilon \equiv \epsilon_{a,a} =
1.0$, $\epsilon_{b,b} = 2.0\epsilon$ and $\epsilon_{a,b} = 1.4\epsilon$,
$d \equiv d_{a,a} = 1.0$, $d_{b,b} = 1.2d$ and $d_{a,b} = 1.1d$ and
$k_{a,a} = k_{b,b} = k_{a,b} = 6/d$. Furthermore, colloidal particles
of both types have same mass: $m_{a} = m_{b} = 1.0$. The potential
is truncated at the cut-off distance $r_c^{a,b}$, where $U_{a,b}(r_c^{a,b})
= 10^{-7} \epsilon_{a,b}$, i.e. negligibly small. The investigations
are done for the particle density of $\rho_0 = 0.675 m_{a}/d_{a,a}^3$.
The choice of such parameters ensures that at densities $\rho_0$
and in the range of considered temperatures, neither crystallization
nor demixing occurs. Four different sample sizes, with $N = 1600$,
$8000$, $17576$ and $32768$ number of particles in a cubic box were
investigated.

In all the simulations, the ambient temperature  is maintained via the 
Dissipative Particle Dynamics (DPD) thermostat \cite{Soddemann2003}. 
The system dynamics is then evolving according to the DPD equation of motion
\begin{equation}
\dot{\vec{r}}_i = \frac{\vec{p}_i}{m_i};  \qquad \dot{\vec{p}}_i = \sum_{j (\neq i)}^N[\vec{F}_{i,j}+\vec{F}_{i,j}^D+\vec{F}_{i,j}^R]\;,
\label{eqn1}
\end{equation}
where $\dot{\vec{r}}_i$ and $\vec{p}_i$ are respectively the position and momentum 
of the particle $i$ with mass $m_i = 1$, $\vec{F}_{i,j} = -\nabla(U_{i,j})$ is the force derivable 
by our Yukawa interaction potential, $\vec{F}_{i,j}^D$ is a dissipative force and $\vec{F}_{i,j}^R$ a 
random force. The dissipative force, $\vec{F}_{i,j}^D$, is proportional to the velocity difference, 
$\vec{v}_{i,j} = \vec{v}_{i}-\vec{v}_{j}$ of particles $i$ and $j$, and it slows down the relative 
motion through a viscous effect controlled by the friction coefficient $\xi$:
\begin{equation} \label{eq:DissapativeForce}
\vec{F}_{i,j}^D = -\xi w^2(r_{i,j})(\hat{\vec{r}}_{i,j} \cdot \vec{v}_{i,j}) \hat{\vec{r}}_{i,j} \;. 
\end{equation}

In the above equation, $\hat{\vec{r}}_{i,j}$ and $r_{i,j}$ are the unitary vector and the 
distance $|\vec{r}_i-\vec{r}_j|$ or module of the vector $\vec{r}_{i,j} = \vec{r}_{i}-\vec{r}_{j}$, 
respectively. The weight function $w(r_{i,j}) = \sqrt{1-r_{i,j}/r_c}$ for $r < r_c^{DPD} = 1.25d$ 
or $w(r_{i,j}) = 0$ otherwise. The relative velocities included in the DPD equations of motion 
ensure the correct thermostatting of the system because of its ability to cancel the drifting
velocity introduced by the shearing process and moreover this choice ensures 
Galilean-invariance and locally conservation of momenta. Concerning the role 
of the friction coefficient in the transient dynamics, it has been shown in previous 
works that there exists a range of values that does not effect neither the transient 
nor the microscopic dynamics \cite{Zausch2009, Chaudhuri2013}, which we confirm 
also for our data in the following section. The random force
\begin{equation}
\vec{F}_{i,j}^R = \sqrt{2k_BT\xi}w(r_{i,j})\theta_{ij} \hat{\vec{r}}_{i,j}\;.
\end{equation}
is implemented in the usual way with $\theta_{i,j} = \theta_{j,i}$ random Gaussian 
variables and a choice of the amplitude $\sqrt{2k_BT\xi}w(r_{i,j})$ obeying the fluctuation-dissipation relation \cite{Groot1997}. 

We use the following protocol for preparing the glassy states,
before applying the external shear stress: Initially, the colloidal
liquid is equilibrated at $T = 0.20$ within standard periodic
boundary conditions. We sample $m$ independent equilibrium states
within this NVT ensemble. Subsequently, all these $m$ liquid samples
are instantaneously quenched to a glassy temperature of $T = 0.05$,
which is well below the mode coupling critical temperature ($T_c =
0.14$) for this binary mixture. Subsequently the samples are aged
for a time $t_w=10^4\tau$. These aged quiescent states are then
subjected to the external shear stress, to probe the response as
explained in the following paragraph. In this work all sample
averages are performed over $m= 80$ samples for the smallest size
studied $N= 1600$ and $m=40$ samples for the larger sizes $N =
8000$, $17576$, $32768$.

For the numerical study of the bulk response of our samples to an applied step 
in shear stress, we employ a novel shear protocol inspired by the work of 
Vezirov et al.~\cite{Vezirov2015a}, whereby a constant macroscopic shear 
stress $\sigma_0$ is maintained via a feedback control scheme, in a system with periodic boundary conditions. 
The feedback is implemented via an evolution equation for the macroscopic shear rate $\dot{\gamma}$:
\begin{equation}
\frac{d \dot\gamma(t)} {dt} = B[\sigma_0 - \sigma_{xy}(t)]\;,
\label{rate}
\end{equation}
where $\sigma_{xy}$ is the macroscopic shear stress, which is being maintained at a value $\sigma_0$.
We apply the shear along the $x$-direction, within the $xy$-plane. The adjustable damping parameter 
$B$ is proportional to the inverse of the timescale for the bulk shear stress to relax to the imposed value. 
We need to choose B appropriately, such that the corresponding time-scale is smaller than any other 
involved time-scale to capture well the correct long time dynamics of the creep and the fluidisation of 
the material; the variations in response with changing $B$ will be discussed below. In the simulations 
we track positions and velocities of constituent particles and the shear stress is measured via the Irving-Kirkwood expression:
\begin{equation}
\sigma_{xy}(t) = \frac{1}{V} \Big \langle \sum_{i}^N \Big [m_i(v_{i,x}(t) v_{j,x}(t)) + \sum_{j > i}^N r_{ij,x}(t) F_{ij,y}(t)\Big ] \Big \rangle\;. 
\end{equation}
The evolution of the system to applied stress is studied by simultaneously integrating, numerically,
the equations of motion for the macroscopic shear rate given in Eqn.~(\ref{rate}) and that of the individual particles given in Eqn.~(\ref{eqn1}). 
The coupling of the two equations ensures that both the motion of the constituent particles as well as the macroscopic deformation
simultaneously adjust to maintain the imposed stress $\sigma_0$. Thus, along with the positions and velocities of the particles,
the macroscopic shear rate, $\dot{\gamma}(t)$, and thereby the strain $\gamma(t)$, are dynamical relevant observables.

All the computations have been carried out using the open source
molecular dynamics program from Sandia National Laboratories, LAMMPS,
\cite{plimpton1995fast}, wherein we have implemented the scheme for
integrating Eqn.~(\ref{rate}). As in previous studies \cite{Zausch2009,
Chaudhuri2013}, we use a time step of $\delta t = 0.0083\tau$, where
$\tau = \sqrt{md^2/\epsilon}$ is our unit of time.

\section{Robustness of the novel protocol}

\begin{figure}[t!]
     \begin{center}
	      \includegraphics[width=\columnwidth]{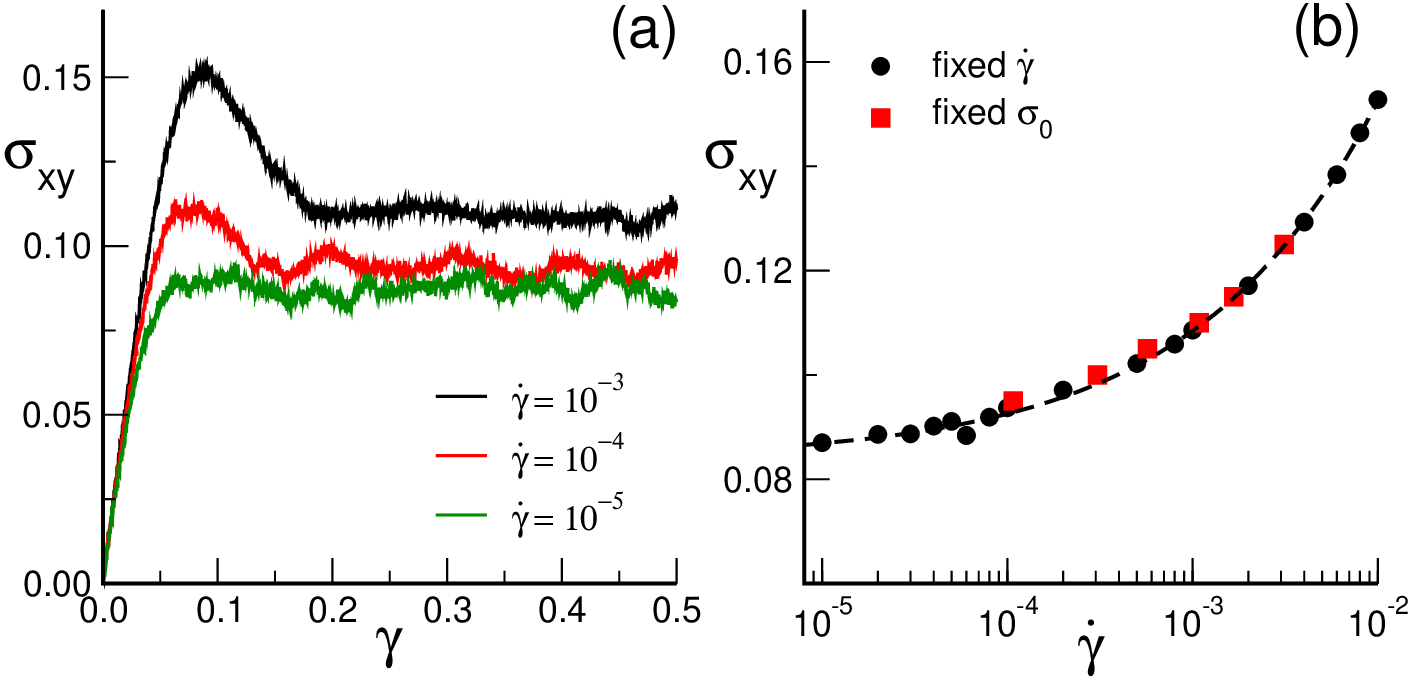}
     \end{center}
    \caption{(a) Shear stress $\sigma_{xy}$ versus shear strain $\gamma$ in a fixed shear rate protocol for the system with $N=17576$ particles, shown for applied shear rates of $10^{-3}, 10^{-4}, 10^{-5}$. (b) Stationary value of the shear stress $\sigma$ as a function of the applied shear rate $\dot{\gamma}$ in the fixed shear rate protocol (black circles), fitted with the Herschel-Bulkley function $\sigma_{xy}=\sigma_y+A\dot{\gamma}^n$, with $\sigma_y=0.0836$, $A=0.534$, $n=0.444$ (dashed line). Superimposed is the data of the steady-state shear rate obtained from the imposed stress protocol imposing $\sigma_0$ (red squares). }
   \label{fig:figure1}
\end{figure}

To ascertain the range of stresses for which the rheology of the
glass needs to be probed, we first reproduce the steady state flow
response of the model at $\rho_0=0.675,T=0.05$ obtained by imposing
a usual shear protocol at imposed constant shear rate with periodic
boundary conditions as shown in Fig.~\ref{fig:figure1}. It is
well-known, that upon the application of a constant shear rate, the
dense amorphous system initially responds macroscopically as an
almost perfect elastic solid (with an elastic shear modulus $G_0$)
before the macroscopic plasticity sets in. The load curve typically
shows a stress overshoot, that will depend on the initial condition
(age) of the quiescent material and the driving rate \cite{varnik2004study,
shrivastav2016heterogeneous, vasisht2017emergence}, before entering
a stationary state, where the macroscopic stress fluctuates around
its steady state stress value $\sigma$. When plotted against the
applied shear rate $\dot{\gamma}$ the dependence of this steady
state shear stress is well described by a phenomenological generalized
Herschel-Bulkley type fit $\sigma=\sigma_y+A\dot{\gamma}^n$. Thus,
the flow curve provides us the range of values, relative to the
yielding threshold, that one can explore to study the response to
the externally applied stress.

\begin{figure}[t!]
     \begin{center}
	      \includegraphics[width=\columnwidth]{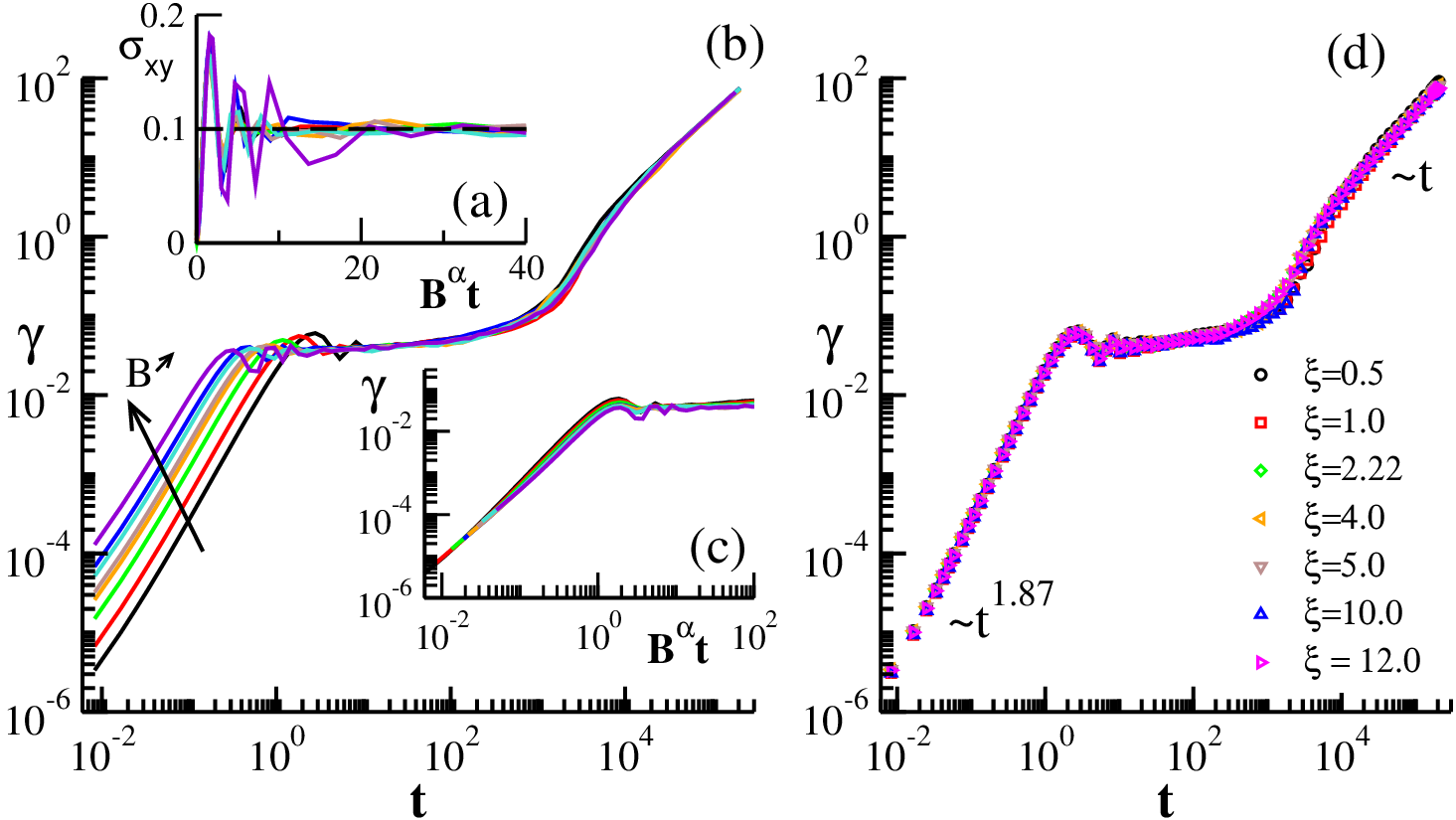}
     \end{center}
\caption{For a system of $N=17576$ particles, (a) evolution of bulk stress $\sigma_{xy}$ for various damping parameters $B$ (0.5, 1.0, 2.22, 4.0, 5.0, 8.0, 10.0, 20.0) in the shear rate evolution equation (Eqn.~\ref{rate}), under an applied step stress $\sigma_0=0.1$ at $t=0$. (b) Corresponding strain evolution $\gamma(t)$ for the different values of $B$. (c) Zoom of the early time dynamics versus a rescaled time $B^{\alpha}t$, with $\alpha\approx0.6$. (d) Effect of the friction coefficient $\xi$ on the strain evolution $\gamma(t)$ for the system with $N=17576$ colloidal particles under imposed stress $\sigma_0 = 0.1$ and damping factor $B=0.5$.}
   \label{fig:figure2}
\end{figure}

To test the robustness of our protocol against the various control
parameters, we consider now one such stress value within that range,
viz. $\sigma_0$ = 0.1, to study the dynamics using the feedback
protocol that we have outlined earlier (Eqn.\ref{rate}). At first,
we check how the choice of the the damping parameter $B$ in
Eqn.\ref{rate} influences the observed behaviour. For a quiescent
glass, i.e. when there in no external shear, the time-averaged shear
stress $\sigma_{xy}$ is zero. At time $t=0$, when the external shear
stress is suddenly set to a finite value, the bulk shear stress
$\sigma_{xy}$ of the material starts to adjust to the externally
imposed one. This evolution is illustrated in Fig.~\ref{fig:figure2}(a)
for various values of the damping parameter $B$. As expected, the
long time bulk shear stress equals the applied magnitude, indicating
that the protocol is working fine. At early times, we observe that
$\sigma_{xy}(t)$ initially increases, responding to the external
forcing, and then oscillates as the feedback mechanism kicks in
before settling down to $\sigma_0$. The magnitude and lifetime of
the observed oscillations is naturally determined by our control
parameter $B$.

The corresponding evolution of the bulk shear strain of the system
$\gamma(t)$ is shown in Fig.~\ref{fig:figure2}(b); each curve is
averaged over $10$ independent trajectories, starting from different
initial glassy configurations, as explained in the model section.
Firstly, the overall behaviour of $\gamma(t)$, as obtained via the
feedback control, is similar to what is observed in experiments
\cite{siebenburger2012creep}, thus validating the numerical scheme.
We also observe that the damping factor $B$ has an influence only
at short-times ($t < 10$) till around the occurrence of the transient
overshoot in $\gamma(t)$, with the magnitude of $B$ determining
the location and height of the overshoot. The larger the value of
$B$, the earlier the system reaches the overshoot and the subsequent
creep regime, which is consistent with the consideration that the
damping factor is inverse to the resistance of the material at the
given deformation. On the other hand, the choice of $B$ does not
have any effect on the response at later times ($t>10$). We see
that the regime where the deformation $\gamma(t)$ shows a plateau,
indicative of the elastic response of the material
\cite{siebenburger2012creep}, depends only on the stress imposed,
that is in fact the same for each curve in Fig.~\ref{fig:figure2}(b).
Similarly, the late time steady state regime, $\gamma(t) \sim t$,
and its onset is also not altered by the magnitude of $B$. In
Fig.~\ref{fig:figure2}(c), we show the time-scaled shear strain for
the different $B$ factors.  We observe that the data collapse for
the early time dynamics for a rescaling of time with a factor
proportional to  $B^{\alpha}$ with an empirical scaling exponent
$\alpha\approx 0.6$. This rescaling also shows that the initial
slope for $t<1$ is the same for all the curves with dynamics that
are close to ballistic motion with $\gamma(t)\sim t^{1.87}$. Note
that the data for the early growth of stress in the system, can
also be collapsed, using the same time rescaling, as shown in
Fig.\ref{fig:figure2}(a). These short-time damped oscillations
result from a competition of inertia of the stress regulation and
the systems elastic properties, an effect that is also commonly
observed in experiments where this ringing effect emerges from a
coupling of the instrument inertia and the elasticity of the material
\cite{baravian1998using, ewoldt2007creep, christopoulou2009ageing}.

To be complete, we also check whether the DPD friction coefficient
$\xi$ has any influence in the strain evolution. The response for
various values of $\xi$ at a given target stress  $\sigma_0$ and
damping factor $B$ is show in Fig.~\ref{fig:figure2}(d). We see
that the dynamics appear insensitive in the range of values of $\xi$
examined: both fluidization time and the creep regime are identical
for the six different friction coefficients investigated. Further
we also checked that the empirical scaling exponent of $B$ does not
depend on the specific value of the friction coefficient $\xi$ (data
not shown here). In all the subsequent data analysis of the following
sections we fixed $B=1$ for the damping factor and $\xi=12$
\cite{Zausch2009} for the friction coefficient, knowing that all
the long time results will not crucially depend on these values.

\section{Creep response}

\begin{figure}[t!]
\begin{center}	
\includegraphics[width=	\columnwidth]{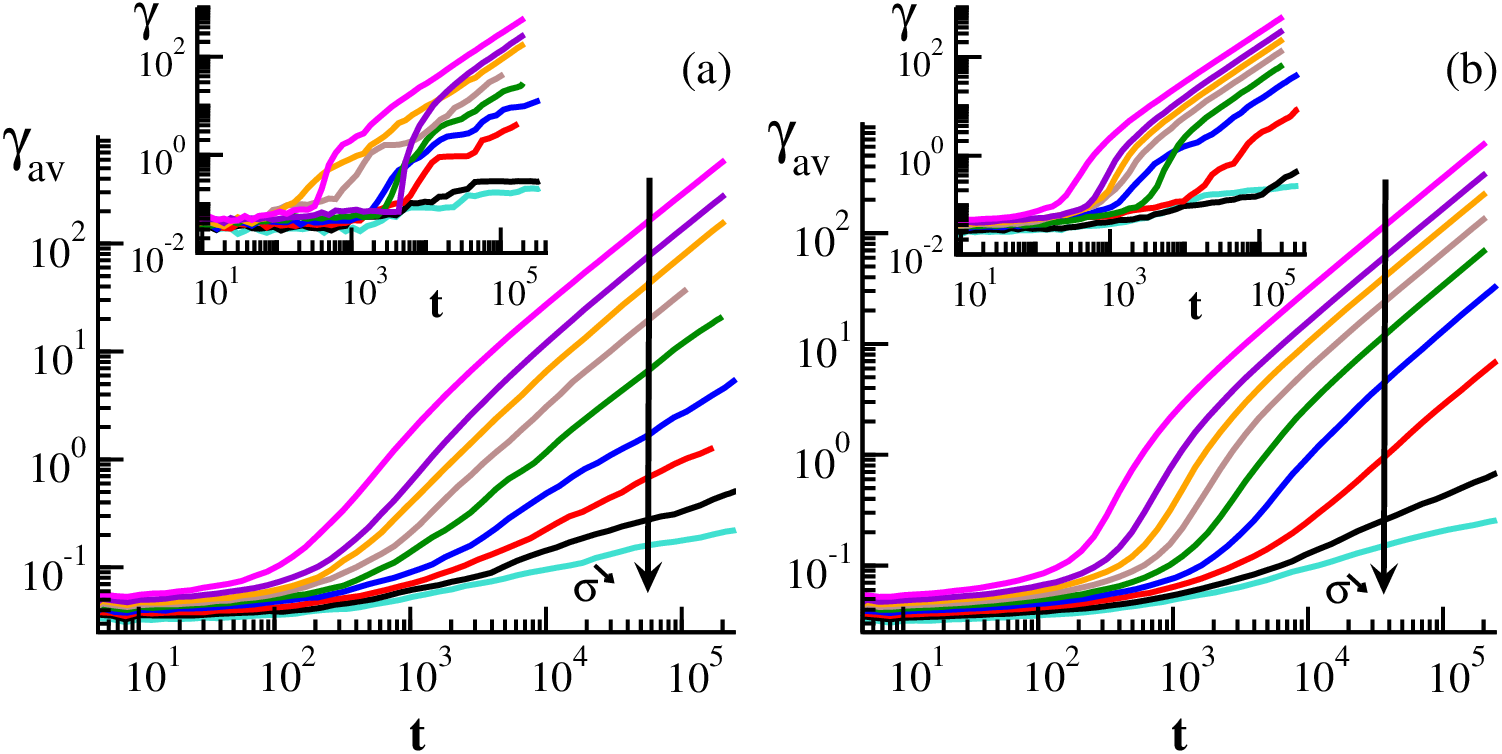}
  \end{center}
\caption{Strain evolution $\gamma(t)$ for a range of imposed stresses $\sigma_0$ (0.080, 0.085, 0.090, $\ldots$, 0.125), 
each averaged over an number of different initial conditions for (a) the smallest system ($N=1600$) and (b) the largest system ($N=17576$) studied. The inset shows typical curves for corresponding individual runs.}
\label{fig:figure3}
\end{figure}

We now systematically discuss the response of the model system, for
different magnitudes of the imposed target stress $\sigma_0$.
Fig.~\ref{fig:figure3} shows the evolution of the strain response
$\gamma_{av}=\langle\gamma(t)\rangle$ for a range of stresses for
a system with (a) $N=1600$ particles, averaged over the ensemble
of $m=80$ initial states, and (b) $N=17576$ particles, averaged
over the ensemble of $m=40$ initial states. In the inset of either
figure, we show some of the individual trajectories for the strain
response $\gamma(t)$, obtained from one specific initial state
within the ensemble for the different applied stresses. First
focusing on the ensemble-averaged $\gamma_{av}$, we observe that
for $\sigma_0 > 0.095$, we observe the long-time steady state regime
$\gamma_{av}(t) \sim t$. For smaller applied stresses, at these
long times ($t>10^4$), the material creeps, i.e. $\gamma_{av}(t)
\sim t^\beta$ ($\beta < 1$), till the longest accessible time-scales.

If one focuses on the macro-response of a single system, as shown
in the insets, we can discern a difference in response. The strain
curves for the smaller system ($N=1600$) demonstrate sharper onset
of flow, compared to the larger system ($N=17576$), indicating
already the existence of finite size effects in the fluidisation
process. Further, signatures of stick-slip motion \cite{varnik2004study}
are also visible in the smaller system, i.e.~there are time windows
over which the deformation remains nearly constants, interspersed
with intermittent jumps. This becomes more prominent as one goes
to smaller $\sigma_0$, i.e.~in the approach to the yield stress
threshold, especially where the long time  linear regime in the
flow response (i.e. $\gamma(t) \sim t$) is not yet visible. Such
intermittency is less visible in the data, for the larger system,
due to the spatial average of response over many smaller intrinsic
blocks.  Further, the stick-slip behaviour is averaged out, while
calculating ensemble averages. Note that such intermittent behaviour
is consistent with the idea of strongly correlated avalanche dynamics
occurring close to the yielding threshold.

\begin{figure}[t!]
\begin{center}
            \includegraphics[scale=0.42]{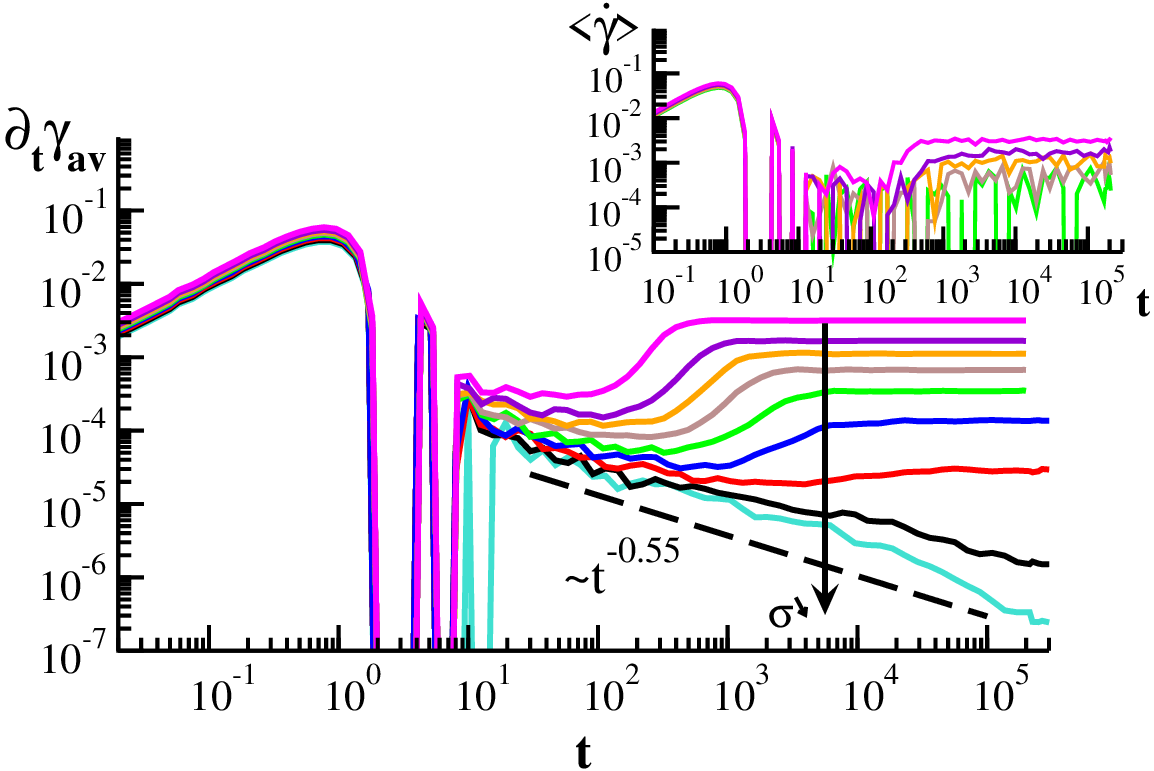}\\  
\end{center}
\caption{The time derivative of the averaged strain response $\partial_t{\gamma}_{\rm av}$ as a function of time, obtained from the data shown in Fig.\ref{fig:figure3}(b) for the same range of stresses. The inset shows the time evolution of the average shear rate that we obtain directly from our protocol (see Eqn.~\ref{rate}), $\langle\dot{\gamma}(t)\rangle$, averaged over the ensemble of 40 initial states, for $N=17576$ and $\sigma_0=0.115,0.110,0.105,0.100$. The data of smaller stresses is not shown here due to the strong fluctuations despite averaging.}
\label{fig:figure4}
\end{figure}

Additionally, in Fig.~\ref{fig:figure4}, we show the corresponding
behaviour of the shear rate $\dot{\gamma}(t)$, computed in two
different ways.  In the inset of Fig.~\ref{fig:figure4}, we  show
the time evolution of the ensemble-averaged $\langle\dot{\gamma}\rangle$
as directly measured from our numerical integration, for a range
of applied stresses. For the larger values of $\sigma_0$,  a long-term
steady state is visible, i.e.  $\langle\dot{\gamma}\rangle$  fluctuates
around a constant value. However, below a certain $\sigma_0$, the
data becomes too noisy, for the averaging done over the ensemble
of $m=40$ initial states. On the other hand, if one considers the
numerical derivative of the ensemble-averaged strain
$\langle\gamma(t)\rangle$ (from Fig.\ref{fig:figure3}(b)), as is
usually done in experiments \cite{Divoux2011, siebenburger2012creep},
we are able to discern the response $\partial_t{\gamma}_{\rm av}$
at all the values of $\sigma_0$  with reasonable precision, as shown
in the main figure of Fig.~\ref{fig:figure4}. For $\sigma_0$ below
a certain value, $\partial_t{\gamma}_{av}$ is not constant at long
times, i.e. there is no observed steady-state regime. Rather, we
observe that $\partial_t{\gamma}_{av}$ decreases with time in a
power law fashion, $\partial_t{\gamma}_{av} \sim t^{\beta-1}$; i.e.
the material is creeping as expected.

To counter-check our results, we gather the data for the measured
steady-state $\dot{\gamma}$ for the different applied $\sigma_0$
(for which a steady state can be reached) to compare with the
shear-rate imposed data. As we can see in Fig.\ref{fig:figure1}(b),
the data from the two different protocols, applied fixed shear rate
and applied fixed stress, are statistically consistent, as it should
be, thereby, confirming again the validity of the numerical scheme
used to impose a fixed target shear stress.

\section{Precursors of fluidisation: finite size effects} 

\begin{figure}[t!]
\includegraphics[width=	\columnwidth]{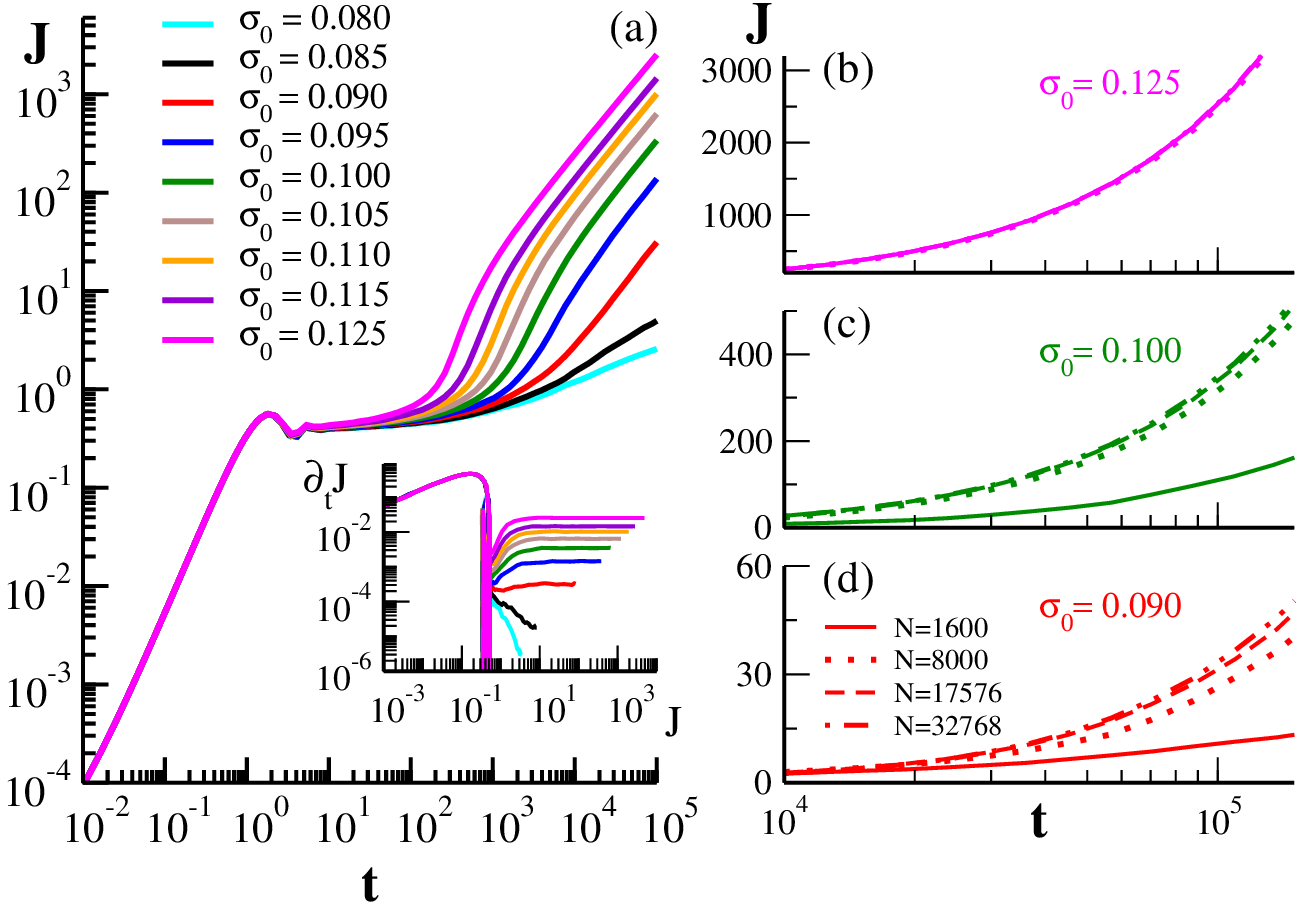}
\caption{(a) Evolution of compliance $J(t)=\gamma(t)/\sigma_0$ for the system with with $N=17576$ particles,
for the range of imposed stresses explored. The inset shows the time derivative of the compliance $\partial_t J$ as a function of the compliance $J$. (b) The compliance $J(t)$ for $N=1600, 8000, 17576, 32768$, for $\sigma_0=0.090, 0.100, 0.125$. 
}
\label{fig:figure5}
\end{figure}

Another quantity that is particularly useful to study the onset of
plasticity in the time-dependent mechanical response is the compliance
$J(t)=\gamma(t)/\sigma_0$\cite{nijenhuis2007non}. In the left panel
of Fig.~\ref{fig:figure5},  for various imposed stresses $\sigma_0$,
the data for $J(t)$ is shown for  the system size of $N=17576$
particles, from which, different features can be observed. For all
the curves,  there is a common region at early times
($t\sim[10^{-2}-10^1]$), where all the data superimpose for the
entire range of applied $\sigma_0$ that we have studied. This is
the regime of elastic deformation and it depends only on the material
itself (in our case, the interactions and/or the dissipation
processes). At later times the curves start to separate out, at
time-scales which increase with decreasing $\sigma_0$, due to the
dependence of the plastic response on the magnitude of imposed
stress. The main plot nicely shows the superposition of all the
compliance curves at short times, where the dynamics remains
completely reversible,  after which deviations start occurring,
with a strong dependence on the imposed target stress $\sigma_0$.
To make the separation of the different curves better visible, we
also plot the time derivative of the compliance as a function of
the compliance itself as shown in the inset of the left panel of
Fig.~\ref{fig:figure5}. The strong drop of the time derivative of
the compliance at a value of the compliance of $J\approx0.5$  marks
the onset of plasticity, beyond which, all the curves start to
depart from the regime of elastic response. Thus, in such a manner,
we are able to delineate the elastic and plastic regimes of the
response, after the application of the external shear stress.

Thereafter, we check how the size of the system affects the compliance.
This is shown on the right panel of  Fig.~\ref{fig:figure5}, for
three different values of applied stress, viz. $\sigma_0=0.125,
0.100, 0.090$. For the largest applied stress, there is no difference
in the response across system sizes. As we go to lower stresses
($\sigma_0=0.100$), the difference starts to be visible, with the
compliance curve for the smallest system ($N=1600$) displaying a
slower increase. The difference across the other sizes becomes more
prominent when one goes even lower in stress ($\sigma_0=0.09$).
Thus, one can conclude that finite size effects are present in the
macroscopic response and this effect becomes more significant as
we approach the yield stress threshold.

\begin{figure}[t!]  
\includegraphics[width=	\columnwidth]{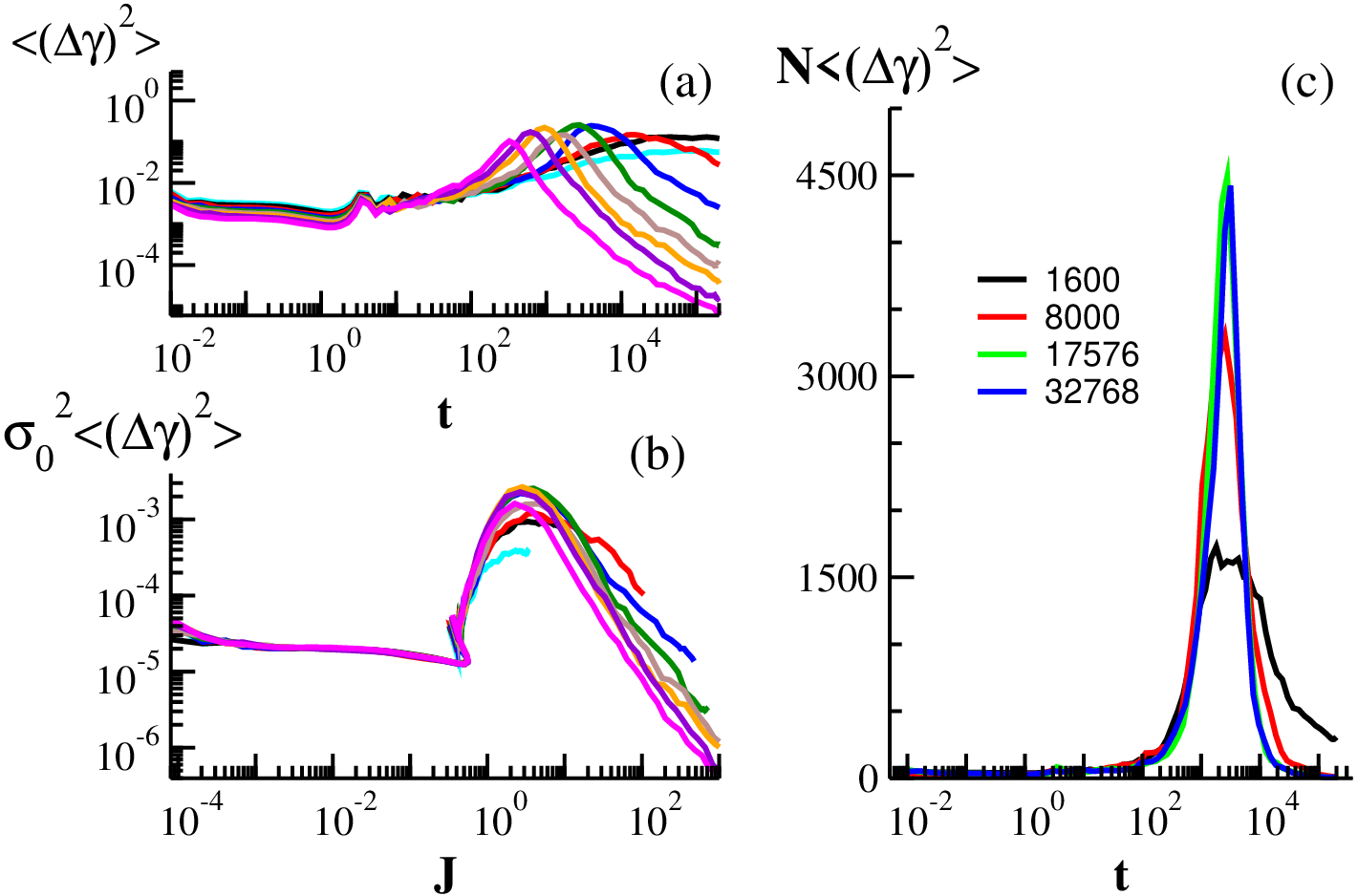}
    \caption{(a) Time evolution of  sample to sample fluctuations in strain $\langle[\Delta\gamma^2(t)]\rangle$, for different applied $\sigma_0$, for $N=17576$. 
(same as in Fig.\ref{fig:figure5})
(b) Corresponding rescaled fluctuations $\langle[\Delta\gamma^2(t)]\rangle\sigma_0^2$ as a function of the compliance $J=\langle\gamma\rangle/\sigma_0$. (c) System size rescaled sample to sample fluctuations in strain $\langle[\Delta\gamma^2(t)]\rangle$ as a function of time for different system sizes $N$.}
   \label{fig:figure6}
\end{figure}

To identify possible precursors of fluidisation, we consider now
the sample-to-sample variations in the mechanical response within
the ensemble of initial states \cite{Chaudhuri2013}.  This is quantified by measuring
the the time-dependent fluctuations in strain, $\Delta\gamma_i(t)$,
of the configuration $i$ in the ensemble, relative to the
ensemble-averaged strain curve $\langle\gamma(t)\rangle$:
\begin{equation}
\Delta\gamma_i^2(t) = \frac{[\gamma_i(t) - \langle\gamma(t)\rangle]^2}{\langle\gamma(t)\rangle^2}
\end{equation}
where $\gamma_i(t)$ is the strain evolution curve for the configuration
$i$. Similar measurements have been employed in experiments, albeit
considering fluctuations across different regions in the sample
\cite{rosti2010fluctuations}. In Fig.\ref{fig:figure6}(a), we show
how the corresponding ensemble-averaged fluctuations
$\langle[\Delta\gamma^2(t)]\rangle$, behave with varying $\sigma_0$,
measured over $m=40$ configurations for the system size $N=17576$.
As we can see, the function is typically non-monotonic
\cite{Chaudhuri2013}, with very little deviations among samples in
the elastic regime and again negligible when all samples are
fluidised.  In between, there is a maximum whose locations shifts
to longer timescales, with decreasing $\sigma_0$, a trend very
similar to what we discussed for fluidisation timescales in the
above discussion related to measured compliances, $J(t)$, shown in
Fig.\ref{fig:figure5}.

The short time elastic response is reflected as well in the sample
to sample fluctuation of the strain. In Fig.~\ref{fig:figure6}(b),
we plot the rescaled fluctuations, i.e. multiplied by $\sigma_0^2$,
versus the compliance ($J$), and find again a perfect collapse of
the data up to the compliance value $J=0.5$, which is where plasticity
sets in. In Fig.~\ref{fig:figure6}(c), we plot for a given stress
$\sigma_0=0.1$, the strain fluctuations, as a function of time, for
different system sizes; to make better comparison, we plot
$N\langle\Delta\gamma^2\rangle$. We find that the size dependence
in the peak of the strain fluctuations is scaling in a non-trivial
manner with system size. This, again, hints at system size effects
in the fluidisation process, consistent with our earlier analysis
related to the compliance curves.

\begin{figure}[t!]
     \begin{center}
            \includegraphics[width=\columnwidth, clip]{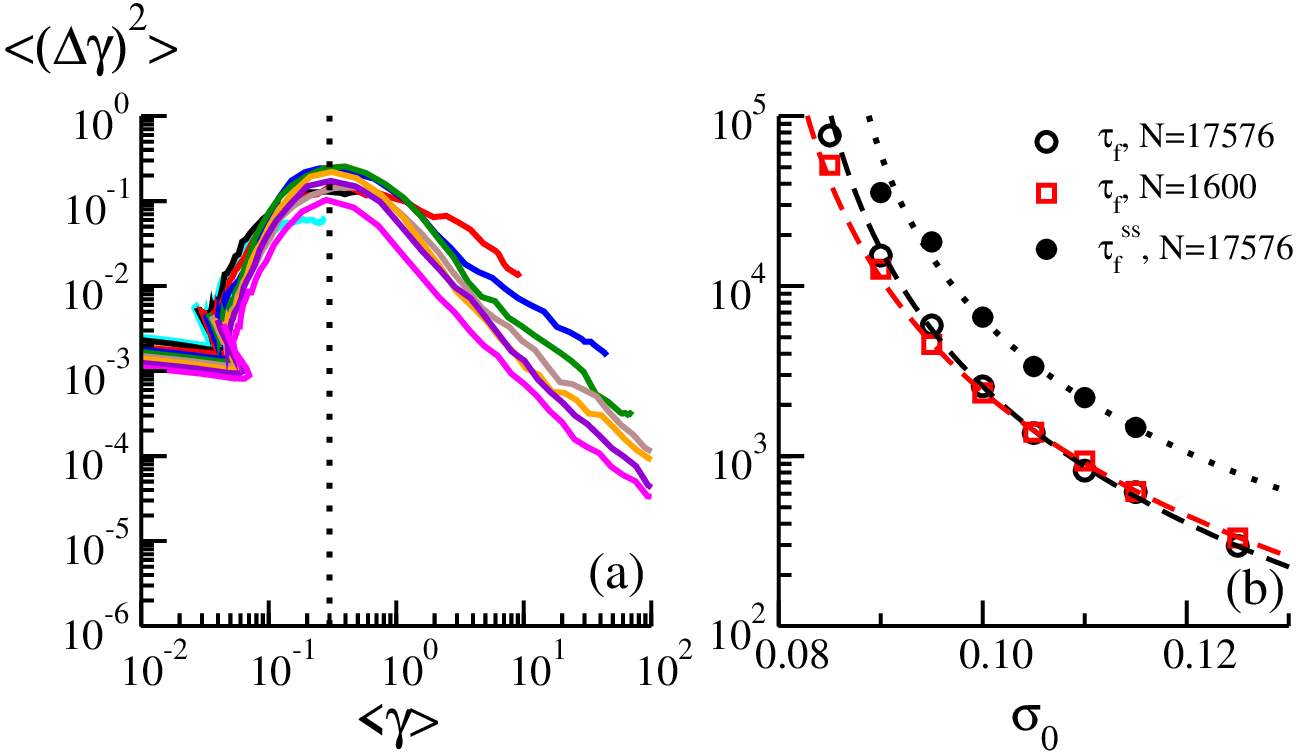}
     \end{center}
    \caption{(a) Sample to sample fluctuations in strain $\langle[\Delta\gamma^2(t)]\rangle$ as a function of strain $\gamma$. (b) Fluidization time $\tau_f$, computed from the peak location in strain fluctuations, as a function of the applied stress ($\sigma_0$) for different system sizes: using open circles for N=17576 and open square for N=1600. The dashed lines show fit using $\sim [\sigma_0-\sigma_s]^{-\zeta}$, obtaining $\sigma_s \approx 0.0772 (N=17576), 0.0801 (N=1600)$, with $\zeta=2.66$. Also shown, using filled circles, is the fluidization time, $\tau_f^{ss}$, obtained from the moment of reaching constant $\partial_t{\gamma}_{av}$, with
the dashed line corresponding to a similar fit as above, with $\sigma_{ss} \approx 0.084$, $\zeta_{ss}=2.18$.
}
   \label{fig:figure7}
\end{figure}

To obtain a rigorous criterion for the fluidisation, we now concentrate
on the late time fluidisation regime.  For this, we study in detail
the location of the peak in the fluctuations, which we interpret
as a precursor to steady state fluidisation. In Fig.~\ref{fig:figure7}(a),
we plot the sample to sample strain fluctuations,
$N\langle\Delta\gamma^2\rangle$, now as a function of strain $\gamma$.
We find that for all imposed stresses, for which the system eventually
shows steady flow,  the maximum of these fluctuations occurs at
approximately the same strain, $\gamma_y \approx 0.3 $, suggesting
a yield strain criterion for the fluidisation process. We use this
yield strain to define in an unambiguous manner the fluidisation
time, $\tau_f$, by monitoring at what time the strain response
$\gamma(t)$ is reaching this value of $\gamma_y$ for any applied
$\sigma_0$.

The results of this procedure are displayed in Fig.~\ref{fig:figure7}(b),
where we plot, using open symbols, the fluidisation time $\tau_f$
as a function of applied stress $\sigma_0$ for different system
sizes. In the literature one finds different propositions for the
dependency of the fluidisation time on the applied external stress,
that can take either an exponential form or a power law divergence
\cite{nicolas2017deformation}. Note that the precision of our data
is not sufficient to discreminate between these two propositions.
Here, we decided to fit the data with a diverging function of the
form $\sim [\sigma_0-\sigma_s]^{-\zeta}$, and we find that the
yielding threshold $\sigma_s$ decreases with system size, viz. we
find the following estimates for the yield stress: $\sigma_s \approx
0.0772 (N=17576), 0.0801 (N=1600)$, using $\zeta=2.66$.

An alternative simple definition for onset of fluidization would
be the time when the onset of steady state flow occurs, i.e. when
$\dot{\gamma}(t)$ becomes constant.  Using data shown in
Fig.\ref{fig:figure4} for $\partial_t{\gamma}_{av}$, we are able
to identify this timescale ($\tau_f^{ss}$) for $N=17576$, and this
too is shown in  Fig.~\ref{fig:figure7}(b), using filled symbols.
In this case, also, a divergence is observed (corresponding fit is
shown in Fig.~\ref{fig:figure7}(b)), and the estimated stress
threshold for onset of flow comes out to be $\sigma_{ss} \approx
0.084$, which is close to the estimated dynamical threshold $\sigma_d
\approx 0.083$, obtaining also an exponent $\zeta_{ss}=2.18$. For
the smaller system, it becomes difficult to have a good quality
measurement, using this definition,  due to lot of fluctuations in
the response.

We note, here, that all such estimates of the threshold or the
exponent for divergence are subject to accuracy in measuring the
timescale, by using any of the above definitions, and this measurement
is strongly subject to proper averaging over a large enough ensemble
of initial states, for each system size.  Nevertheless, our observation
that yielding threshold decreases with increasing system size is
consistent with experimental measurements \cite{weiss2014finite},
and also the hypothesis that within a larger system, there is more
likelihood of finding a weak zone that would lead to earlier yielding,
as compared to a smaller system. This has also been observed for
crystal plasticity \cite{zapperi2012current, ispanovity2013average}.
In any case, more  numerical investigations are needed to clarify
this in the context of yielding of amorphous systems.

\section{Concluding discussion}

In this paper, we have studied several statistical aspects of the
creep response of a model soft glass by means of particle based
simulations. We propose a novel method to impose a target shear
stress to the system, using periodic boundary conditions, which
allows for studying the bulk rheology under applied shear stress,
and we tested the robustness of this approach in details.  This
over-constrtained method for applying the shear combined with our
feedback mechanism allows for the study of the bulk dynamics of a
system undergoing creep without imposing a particular flow profile.
In this way, we are able to address questions like finite size
effects in the dynamics without having to deal with over-constraint
dynamics or unwanted wall effects.

We use this method on a three dimensional binary mixture of particles
interacting via a Yukawa potential to be able to compare to earlier
works, that had been implemented with walls \cite{Chaudhuri2013}.
The resulting mechanical response takes the form of a usual creep
curve with a fast elastic response followed by a power-law decay
of the strain rate, with exponents that lie in the typical range
of experiments and earlier simulations. At longer times, depending
on the stress applied we see a rapid acceleration of the deformation
rate, eventually leading to a steady flow. We show that the
corresponding flow curve for this steady flow agree with the one
from strain rate controlled protocols, thus validating the protocol.

By investigating the compliance curves, we demonstrate how to
evidence the onset of plasticity that occurs above a given value
of the compliance, independent of the applied stress and system
size. The subsequent plastic regime exhibits finite size effects.
In this regime, the dependence on the system size becomes stronger
as we lower the applied stress value,  approaching the yielding
transition.

Next, we demonstrate that the  sample-to-sample fluctuations of the
strain curves show a strong non-monotonic behavior and we identify
the  prominent peak observed in these fluctuations as a precursor
for fluidisation. Since the position of this peak as a function of
strain appears neither stress  nor system size dependent, we use
this feature to unambiguously define a fluidisation time that we
measure as a function of imposed stress and system size.

The analysis of the fluidisation time dependence on the imposed
external shear stress suggests a divergence of the time to reach a
flowing state close to the yield threshold. Although more statistics
will be necessary to be more conclusive and obtain quantitative
relations, the system size analysis of the fluidisation time seems
to indicate that smaller systems exhibit a longer fluidisation
times. This result is in agreement with the weakest link theory,
that predicts that smaller systems exhibit a larger strength, since
the smaller the sample, the smaller also the probability to encounter
a weak spot, that would lead to failure \cite{zapperi2012current,
ispanovity2013average}.

The above observations, related to size effects,  as well as the
strongly intermittent dynamics, observed in the individual creep
curves for small applied stress and small system size, are likely
related to the development of dynamical heterogeneities, with
increasing spatial scale, close the yield threshold. Such stick-slip
behaviour is also reminiscent of avalanche type dynamics for the
plasticity close to the yielding point \cite{talamali2011avalanches, salerno2012avalanches}.  All these would indicate
strongly correlated dynamics, with decreasing stress,  in tune with
the hypothesis that yielding corresponds to an underlying dynamical
phase transition, accompanied by critical dynamics \cite{lin2014scaling,liu2016driving}. Further extensive
investigations are necessary to test out these ideas and the numerical
protocol,  described here, would be an useful mode for such studies.

\section{Acknowledgements}

K.~M.~and P.~C.~acknowledge financial support from CEFIPRA Grant No. 5604-1 (AMORPHOUS-MULTISCALE). 
K.~M.~and R.~C.~acknowledge financial support under ANR Grant No. ANR-14-CE32-0005 (FAPRES).
Part of the computations presented in this paper were performed using the CIMENT infrastructure (https://ciment.ujf-grenoble.fr), which is supported by the Rh\^one-Alpes region 
(GRANT CPER07\_13 CIRA: http://www.ci-ra.org) and the CURIE cluster at TGCC thanks to the Grand Equipement National de Calcul Intensif Project No.~t201509747. 
We also acknowledge the use of  HPC facilities at IMSc, Chennai, for our computations.
Further we thank Jean-Louis Barrat, Eric Bertin and Luca Cipelletti for interesting discussions on the topic.
%




\balance



\bibliography{creep-paper}
\bibliographystyle{rsc} 

\end{document}